\documentclass[twocolumn,showkeys,aps,prl,showpacs]{revtex4-1}
\UseRawInputEncoding
\usepackage{graphicx}
\usepackage[CJKbookmarks,dvipdfm,colorlinks,linkcolor=blue,citecolor=blue]{hyperref}
\usepackage{amsmath, amssymb}
\usepackage{makecell}
\usepackage{multirow}
\usepackage{CJK}
\usepackage{mathrsfs}
\usepackage{bm}
\usepackage{amsmath}
\usepackage{dcolumn}
\usepackage{epstopdf}
\usepackage{dsfont}
\usepackage{amssymb}
\usepackage{tabularx}
\usepackage{array}
\usepackage{float}
\usepackage{color}
\usepackage{epstopdf}
\usepackage{mathrsfs}
\usepackage{extarrows}

\begin{document}

\title{Valley  polarization  in twisted altermagnetism}
\author{San-Dong Guo$^{\textcolor[rgb]{0.00,0.00,1.00}{\dagger}}$}
\email{sandongyuwang@163.com}
\affiliation{School of Electronic Engineering, Xi'an University of Posts and Telecommunications, Xi'an 710121, China}

\author{Yichen Liu}
\thanks{These authors contributed equally to this work.}
\affiliation{Centre for Quantum Physics, Key Laboratory of Advanced Optoelectronic Quantum Architecture and Measurement (MOE),
School of Physics, Beijing Institute of Technology, Beijing 100081, China}

\author{Cheng-Cheng Liu}
\email{ccliu@bit.edu.cn}
\affiliation{Centre for Quantum Physics, Key Laboratory of Advanced Optoelectronic Quantum Architecture and Measurement (MOE),
School of Physics, Beijing Institute of Technology, Beijing 100081, China}

\begin{abstract}
The combination of altermagnetism, twistronics and valleytronics is of great significance for potential
applications in advanced electronic  devices. Twisted magnetic van der Waals bilayers have been identified as an ideal platform for altermagnetism of any type, such
as $d$-wave, $g$-wave, and $i$-wave, by choosing the constituent monolayer with specific symmetry [\textcolor[rgb]{0.00,0.00,1.00}{arXiv:2404.17146 (2024)}].
 Here, we propose a way for achieving valley  polarization  in twisted altermagnetism by applying out-of-plane external electric field.
 Since the out-of-plane electric field creates a layer-dependent electrostatic potential, the valleys form different layers will stagger,  producing  valley  polarization. We also demonstrate the effectiveness of our
proposed way using the twisted tight-binding model.  It is found that the applied electric field can also induce valley/spin-gapless semiconductor  and half metal  besides valley polarization. Based on first-principles calculations, our proposed way  to achieve valley polarization  can be verified in  twisted bilayer VOBr and monolayer  $\mathrm{Ca(CoN)_2}$ as a special  twisted altermagnet.
These findings provide new opportunities for innovative spintronics, twistronics and valleytronics applications.

\end{abstract}

\maketitle

\section{Introduction}
Due to lacking any net magnetic moment,
the  antiferromagnetic (AFM)  materials are robust to external
magnetic perturbation,  and  have  ultra-high dynamic speed\cite{k1,k2}.
In general, these AFM materials lack spin splitting, which is considered not to be conducive to the generation of spin-polarized currents\cite{k3}.
Recently,  a category of collinear crystal-symmetry compensated magnetic ordering called altermagnetism  has been emerging as an exciting research
landscape\cite{k4,k5,k6,k13}. The altermagnetism can realize  spin-splitting, which is only originated from the simple AFM
ordering of special magnetic space group without the help of relativistic spin-orbital coupling (SOC).
Several bulk and two-dimensional (2D) materials have been predicted to be  altermagnetism\cite{k7,k7-1,k7-2,k8,k9,k10,k10-1,k11,k12}.

\begin{figure}
  \includegraphics[width=8cm]{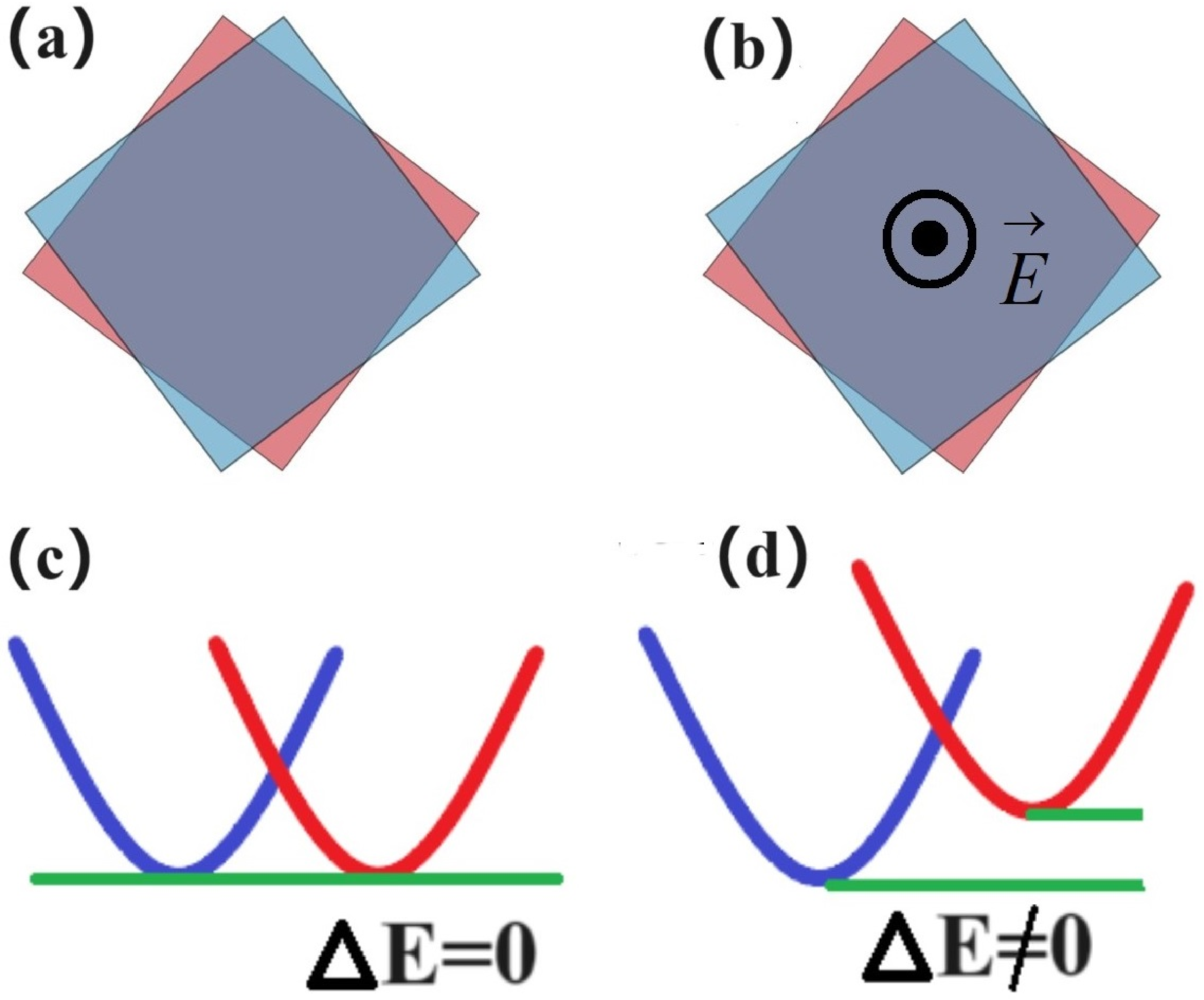}
\caption{(Color online)(a)/(b): a twisted altermagnetic bilayer without/with out-of-plane electric field; (c) and (d): two valleys near the Fermi level are from different layers with opposite spin polarization. For (a) case,  the degeneracy of electron spin is removed, but it lacks valley polarization (c).   For (b) case,  the degeneracy of both spin  and valley are lifted, producing spin-valley polarization (d).  In (b),  the black dotted circle means an out-of-plane electric field. In (c) and (d), the spin-up
and spin-down channels are depicted in blue and red.}\label{sy}
\end{figure}

Twisted-angle 2D systems exhibit novel and tunable properties, such as flat
bands, nontrivial topology, emergent symmetries, enhanced
correlations, and strong electron-phonon coupling,
due to  the charge localization induced by the formation of $\mathrm{moir\acute{e}}$ superlattice\cite{k14,k15,k16,k17,k18,k19}.
 Twisted magnetic systems
 have also been extensively
studied due to their novel properties\cite{k20,k21,k22,k23,k24}, including
magnetic ground states, multiflavor magnetic states, nonrelativistic spin-momentum coupling, noncollinear magnetic states and $\mathrm{moir\acute{e}}$ magnon
bands.
Recently, altermagnetism has been achieved in twisted magnetic van der Waals (vdW) bilayers by introducing   a key
in-plane 2-fold rotational operation, which takes one of all five 2D Bravais lattices\cite{k25}.
The symmetry of spin splitting, i.e.,
$d$-wave, $g$-wave or $i$-wave, can be realized by choosing out-of-plane rotational symmetry.
 Twisted altermagnetism provides  a general,  adjustable and ideal
platform to  explore multifunctional features, for example a combination of altermagnetism, twistronics and valleytronics.

Valleytronics by utilizing valley degree of
freedom to encode and process information provides remarkable opportunities for developing energy-efficient devices \cite{q1,q2,q3,q4}.
In AFM materials especially for altermagnetism,  realizing  valley polarization  is more meaningful for valleytronic application.
Twisted altermagnetism provides a general platform for exploring valley polarization, because it can possess two equivalent valleys from different layers\cite{k25}. Out-of-plane electric field can eliminate the  degeneracy of valleys from different layers\cite{gsd}.
Here, we propose a way to achieve valley polarization  by electric field in twisted altermagnets.

\begin{figure}
  \includegraphics[width=8cm]{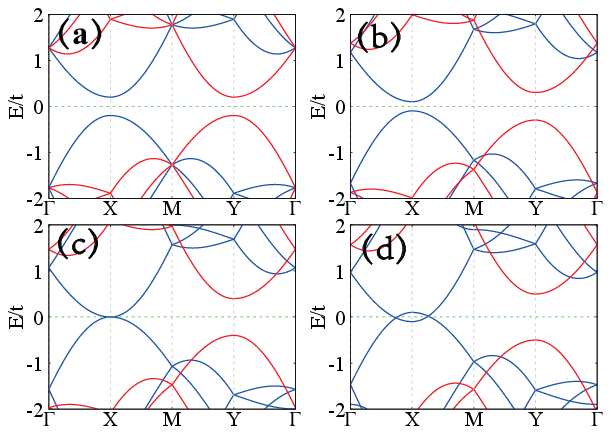}
\caption{(Color online) The energy band structure of the twisted model without (a)/with (b, c, d) electric field by using  $\delta=0.0t$ (a), $0.1t$ (b), $0.2t$ (c), $0.3t$ (d).  The spin-up
and spin-down channels are depicted in blue and red. The other parameters $t_1=-0.5t_2=t$, $M=3.2t$  are chosen. }\label{band}
\end{figure}

\section{Computational detail}
  Within density functional theory (DFT)\cite{1}, the spin-polarized  first-principles calculations are carried out by using the standard VASP package\cite{pv1,pv2,pv3} within the projector augmented-wave (PAW) method. We use  generalized gradient
approximation  of Perdew-Burke-Ernzerhof (PBE-GGA)\cite{pbe}  as the exchange-correlation functional.
The kinetic energy cutoff  of 500 eV,  total energy  convergence criterion of  $10^{-6}$ eV, and  force convergence criterion of 0.001 $\mathrm{eV.{\AA}^{-1}}$ are adopted.
To account for electron correlation of V-3$d$ and  Co-3$d$ orbitals, a Hubbard correction $U_{eff}$=3.0 eV is employed within the
rotationally invariant approach proposed by Dudarev et al\cite{du}.
By taking a vacuum of more than 16 $\mathrm{{\AA}}$, the out-of-plane interaction is neglected.
 A sufficient Monkhorst-Pack k-point meshes are used to sample the Brillouin zone (BZ) for calculating electronic structures.

\section{Approach and  Model}
By  using the in-plane  2-fold rotational operation, stacking a twisted bilayer structure with interlayer AFM ordering can  construct altermagnetism\cite{k25},  because the design principle can produce the compensated collinear magnetic order, and make  the sublattices with opposite spins connect by a rotation symmetry. The method  is a general approach, which takes one of all five 2D Bravais lattices, including  oblique lattice, rectangular/centered-rectangular lattice,  square lattice  and  hexagonal
lattice.

Here,  we propose a way to achieve valley  polarization  in twisted altermagnetism, which is valid for all five 2D Bravais lattices.
Next, a square lattice, as an example, is used  to illustrate our proposal.  A twisted bilayer of  square lattice with interlayer AFM ordering is shown in \autoref{sy} (a), which possesses in-plane  2-fold rotational symmetry.  We assume that the altermagnetic twisted bilayer possesses two degenerate valleys with opposite spin from different layers near the Fermi level (\autoref{sy} (c)). To induce valley polarization, an out-of-plane electric field is applied for the altermagnetic twisted bilayer (\autoref{sy} (b)).  Because  an out-of-plane electric field can create a layer-dependent electrostatic potential, and the electronic bands in different layer will stagger\cite{gsd}, which  gives rise to the valley polarization (\autoref{sy} (d)). It is also readily comprehensible that the transformation of valley polarization can be achieved by reversing the direction of electric field.

\begin{figure*}
  \includegraphics[width=16cm]{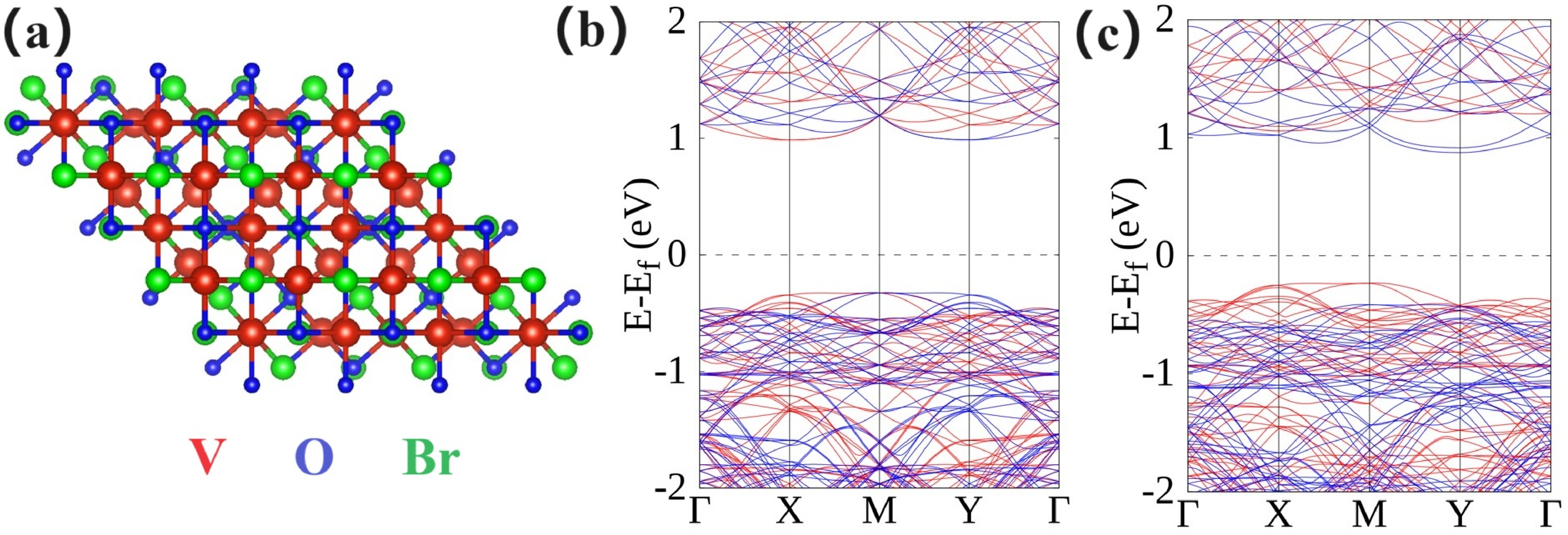}
\caption{(Color online)  The crystal structures (a)  and energy  band structure ($E$=0 $\mathrm{V/{\AA}}$ (b) and 0.02 $\mathrm{V/{\AA}}$ (c)) of twisted bilayer VOBr with the twist angle being 48.16$^{\circ}$.  The spin-up
and spin-down channels are depicted in blue and red.  }\label{v}
\end{figure*}

For a square lattice, the  Hamiltonian of  monolayer model can be  expressed as:
\begin{equation}\label{d-1}
h(\bm k)=t_1\cos(k_x)+t_2\cos(k_y)
\end{equation}
where $t_i$  and $\bm k$ are hopping parameters and wave vector. Firstly, monolayers  are stacked into a bilayer structure. And then, by flipping the upper layer, and
rotating the upper and lower layers by $-\frac{\theta}{2}$ and $\frac{\theta}{2}$,  a $\mathrm{moir\acute{e}}$ supercell can be constructed. The twisted tight-binding (TB) model
can be written as\cite{k25}:
\begin{equation}\label{d-2}
H=\left(\begin{matrix}
H_{11}& T\\
T^\dag  & H_{22}\\
\end{matrix}\right),
\end{equation}

\begin{equation}\label{d-3}
H_{11}=C_{2\parallel}^m h(R(-\frac{\theta}{2})\bm k)C_{2\parallel}^m+M s_z+\delta ,
\end{equation}
\begin{equation}\label{d-4}
H_{22}= h(R(\frac{\theta}{2})\bm k)-M s_z-\delta .
\end{equation}
Where $C_{2\parallel}^m$ and $R$ are the flip operation and rotation matrix;
$T$, $s_z$ and  $M$ represent the interlayer coupling, the spin
on each layer and the magnetic moment; $\delta$=$Ed/2$ with $E$/$d$ being effective electric field intensity/interlayer distance. Here, the interlayer coupling can be neglected.

The energy band structures of our twisted
TB model are plotted in \autoref{band} without/with electric field.
Without electric field, $d$-wave altermagnetism can be observed, and two degenerate valleys with opposite spin from different layers, i.e., spin-layer locking, appear in both conduction  and valence bands near the Fermi level (\autoref{band} (a)). When an out-of-plane electric field is applied, the valley polarization can be achieved (\autoref{band} (b)). When continuously increasing $E$, the  valley/spin-gapless semiconductor (\autoref{band} (c)) and half-metal (\autoref{band} (d)) can also be realized.  For valley/spin gapless semiconductor, both electron and hole can be fully valley/spin polarized\cite{gsd1,gsd2}.  No threshold energy
is required to make electrons from occupied states to
empty states for these gapless electronic states.  The half-metal
possesses full electron spin/valley polarization\cite{k1}.  These electronic states are the most important candidates
for applications as spintronic/valleytronic materials.

\begin{figure}
  \includegraphics[width=8cm]{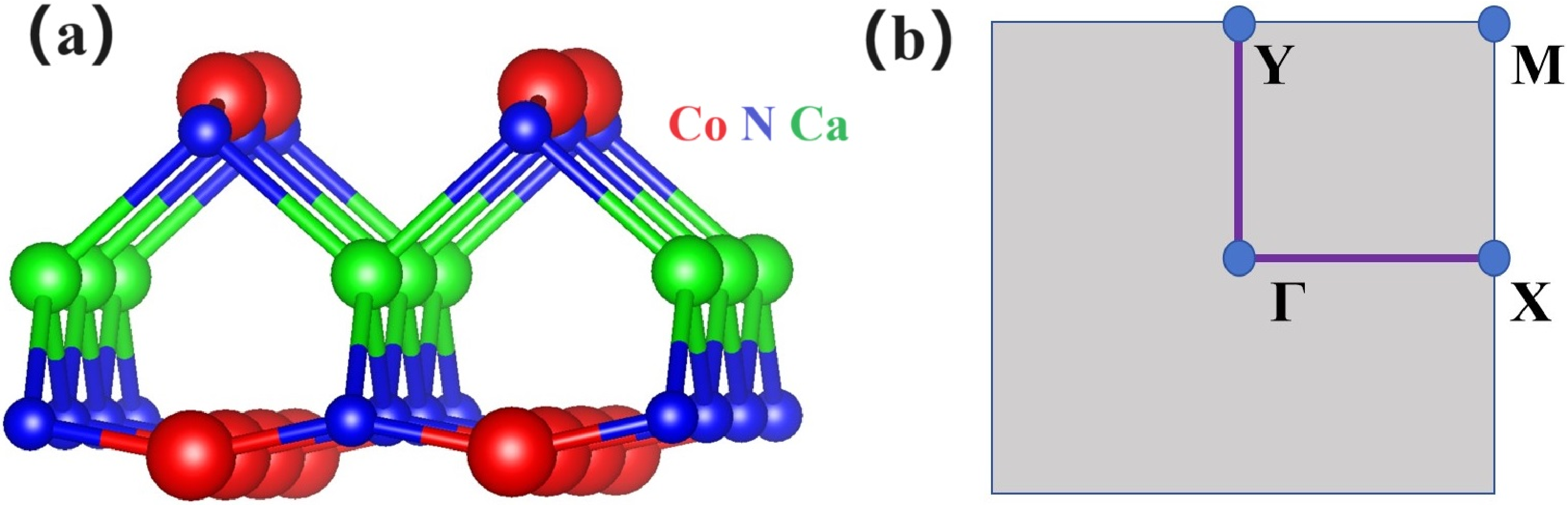}
\caption{(Color online) For  $\mathrm{Ca(CoN)_2}$ monolayer, the crystal structure (a) and the first BZ with high-symmetry points (b).}\label{st}
\end{figure}

\section{Material verification}
The twisted bilayer VOBr with the twist angle being 48.16$^{\circ}$\cite{k25} is used as the first example to verify our proposal.  The experimental lattice constants a=3.775 $\mathrm{{\AA}}$ and b = 3.38 $\mathrm{{\AA}}$\cite{gsd3} are used to construct twisted bilayer VOBr, and  its crystal structures are plotted in \autoref{v} (a). \autoref{v} (b) and (c) show the intrinsic energy band structure of twisted bilayer VOBr and the one with effective electric field $E$=0.02  $\mathrm{V/{\AA}}$  by using GGA+$U$ with $U$=3 eV.
The intrinsic band structures show $d$-wave
altermagnetism, and  there are
two equivalent valleys with  opposite spin from different layers for the conduction  bands.
When applying an out-of-plane electric field of $E$= 0.02 $\mathrm{V/{\AA}}$, an induced  layer-dependent electrostatic potential  gives rising to obvious valley splitting  for conduction bands, producing spin/valley polarization.

In fact, $\mathrm{Ca(CoN)_2}$ monolayer  can be considered as a special altermagnetic twisted bilayer, and its crystal structure  and  first BZ are plotted in \autoref{st}. Monolayer $\mathrm{Ca(CoN)_2}$ has a square lattice structure, whose  unit cell contains five atoms with five-atomic layer sequence of Co-N-Ca-N-Co. The  $\mathrm{Ca(CoN)_2}$ has been predicted to be an altermagnet with A-type AFM ordering, which  is dynamically  and thermodynamically stable\cite{yz}. The optimized lattice constants $a$=$b$=3.548 $\mathrm{{\AA}}$ by GGA+$U$ with $U$=3 eV.

\begin{figure*}
  \includegraphics[width=16cm]{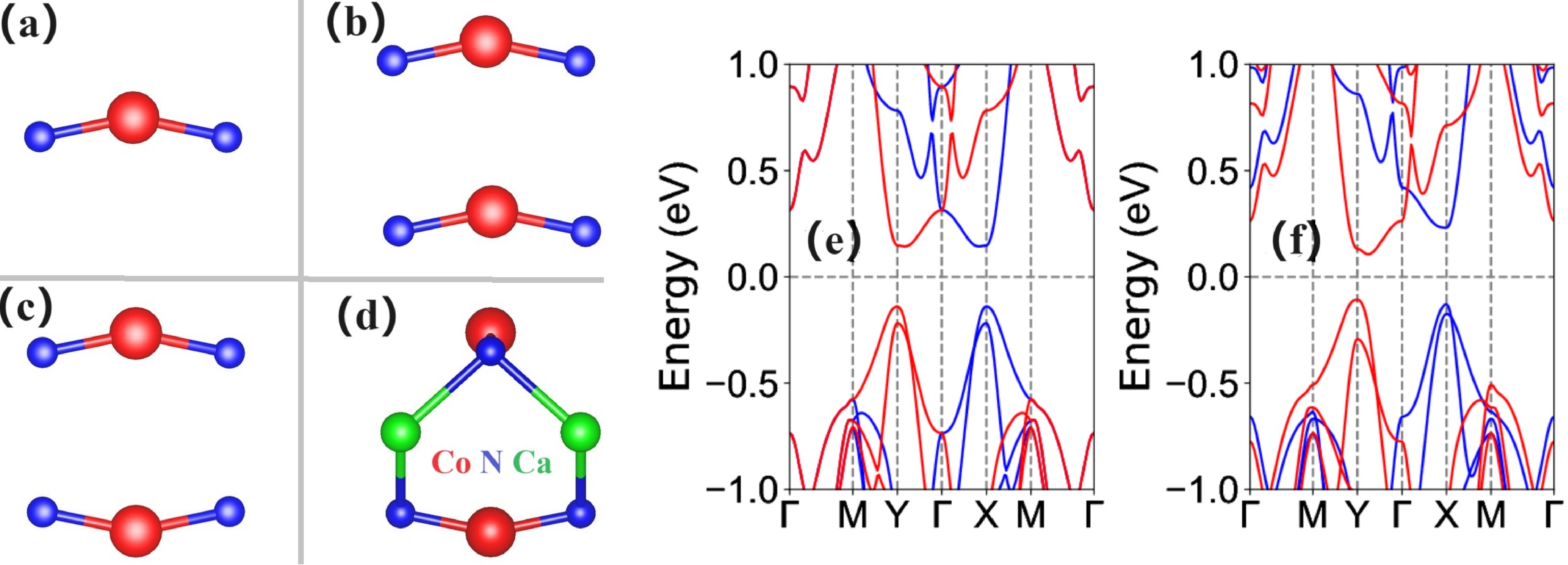}
\caption{(Color online)(a): the building block CoN layer; (b): a bilayer structure  $\mathrm{(CoN)_2}$; (c): the lower  layer of $\mathrm{(CoN)_2}$ is flipped; (d): a 90$^{\circ}$ twist operation, and the two layers are connected by Ca layer; (e) and (f): the energy
band structures of $\mathrm{Ca(CoN)_2}$ with $E$=0 $\mathrm{V/{\AA}}$ (e) and 0.04 $\mathrm{V/{\AA}}$ (f), and the spin-up
and spin-down channels are depicted in blue and red. }\label{st1}
\end{figure*}

Next, we explain why  $\mathrm{Ca(CoN)_2}$ monolayer  is a special altermagnetic twisted bilayer.
The CoN layer can be used as a building block (\autoref{st1} (a)).
Initially, two monolayer CoN are stacked into a bilayer structure that requires AFM ordering (\autoref{st1} (b)). Subsequently, the lower layer CoN is flipped to introduce an in-plane 2-fold rotational symmetry ((\autoref{st1} (c))). Then, the upper and lower layers are rotated by 45$^{\circ}$
 and -45$^{\circ}$, and  the two layers are connected by Ca layer ((\autoref{st1} (d))).
 After the twist, the in-plane 2-fold rotational symmetry still remains, which
connects two layers with opposite spins and plays a key
role in generating altermagnetism.    The build process of $\mathrm{Ca(CoN)_2}$ is in accordance with  the approach of generating altermagnetism by twisting  magnetic vdW bilayers\cite{k25}.

\autoref{st1} (e) and (f) show the the energy
band structures of  $\mathrm{Ca(CoN)_2}$ with $E$=0 $\mathrm{V/{\AA}}$  and 0.04 $\mathrm{V/{\AA}}$ by using GGA+$U$.
At $E$=0 $\mathrm{V/{\AA}}$, there are
two equivalent valleys for both the conduction and valence bands, and they possess  opposite spin from different layers, which provides possibility to induce valley polarization by an out-of-plane electric field. Due to $\mathrm{D_2}$ point group of   $\mathrm{Ca(CoN)_2}$, it belongs to $d$-wave altermagnetism\cite{k25}. The  total magnetic moment of  $\mathrm{Ca(CoN)_2}$  is strictly equal to 0 $\mu_B$, and the magnetic moments of two Co atoms are  2.521 $\mu_B$ and -2.521 $\mu_B$, respectively.
When an out-of-plane electric field of $E$= 0.04 $\mathrm{V/{\AA}}$ is applied, a layer-dependent electrostatic potential is induced, give rising to valley splitting of 125/21 meV for conduction/valence bands.  In addition, the magnitude of magnetic moments of  two Co atoms (2.508 $\mu_B$ and -2.528 $\mu_B$) becomes unequal, although the total magnetic moment of $\mathrm{Ca(CoN)_2}$ is zero.

\section{Conclusion}
In summary,  we propose an alternative strategy to achieve
valley polarization  based on twisted altermagnetism by applying an out-of-plane electric field.
Our proposed way is applicable to
all five 2D Bravais lattices, and the key is that twisted altermagnetism has two equivalent valleys from different layers near the Fermi level.
This ensures that the external electric field causes the two valleys to experience different potential energies, resulting in valley polarization.
It is demonstrated that twisted bilayer VOBr and monolayer  $\mathrm{Ca(CoN)_2}$ as a special  twisted altermagnet  are possible candidates for realizing valley polarization.
 Our works reveal a new 2D family of AFM valleyelectronic  materials, which allow
multifunctional  device applications.

\begin{acknowledgments}
The work is supported by the NSF of China (Grant No. 12374055), the
National Key R\&D Program of China (Grant No. 2020YFA0308800),
and the Science Fund for Creative Research Groups of NSFC (Grant No.
12321004).  We are grateful to Shanxi Supercomputing Center of China, and the calculations were performed on TianHe-2.
\end{acknowledgments}

\end{document}